\documentclass[prb,aps,amssymb,footinbib,showpacs]{revtex4-1}
\usepackage{amssymb}
\usepackage{amsmath}
\usepackage{times}
\usepackage{latexsym}
\usepackage{graphicx}
\usepackage{color}
\usepackage{bm}
\usepackage{subfigure}
\begin{document}

\title{ Geometry, Number Theory and the  Butterfly Spectrum  of Two-Dimensional Bloch Electrons}

\author{Indubala I Satija}

\affiliation{Department of Physics and Astronomy, George Mason University, Fairfax, VA 22030}

\date{\today}

\begin{abstract}
{We take a deeper dive into the geometry and the number theory that underlay the butterfly graphs of the Harper and the generalized Harper models of Bloch electrons in a magnetic field.
Root of the  number theoretical characteristics of the fractal spectrum is traced to a close relationship between the Farey tree -  the hierarchical tree that generates all rationals and the Wannier diagram - a graph that labels all the gaps of the butterfly graph.  The resulting Farey-Wannier hierarchical lattice of trapezoids provides  geometrical representation of the nested pattern of butterflies in the butterfly graph.  Some  features of the energy spectrum such as absence of some of the Wannier trajectories in the butterfly graph fall outside the number theoretical framework,
can be stated as a simple rule of ``minimal violation of  mirror symmetry". In a generalized Harper model, Farey-Wannier representation prevails as the lattice regroups to form some hexagonal unit cells creating new {\it species} of butterflies.}

\end{abstract}

\maketitle 

\section{ Introduction}

The ``butterfly graph" - a quantum fractal, is a graph of energy spectrum of Bloch electrons in a two-dimensional square lattice subjected to a traverse magnetic field.  Resembling a butterfly, it consists of 
self-similar pattern of nested sets of copies of itself. Commonly referred as  the ``Hofstadter butterfly" after its  discovery by Douglas Hofstadter in $1976$\cite{Hof}, the subject has attracted a broad spectrum of physics and mathematics community\cite{book, PT, Math1}. Furthermore, there are various recent attempts to capture this iconic spectrum in various laboratories\cite{Dean}.  The butterfly graph as a whole describes all possible phases - the integer quantum Hall states, of a two-dimensional electron gas\cite{TKNN} that arise as one varies the electron density and the magnetic field.  Each phase is characterized by an integer that represents the quantum number of Hall conductivity. Recent studies\cite{book,Sat16,SW,Sat21}  have described various features of the butterfly spectrum using pure number theoretical reasoning and the quantum fractal is found closely related to  some abstract mathematical sets.\\

\begin{figure}[htbp] 
\includegraphics[width = 0.85\linewidth,height=.7\linewidth]{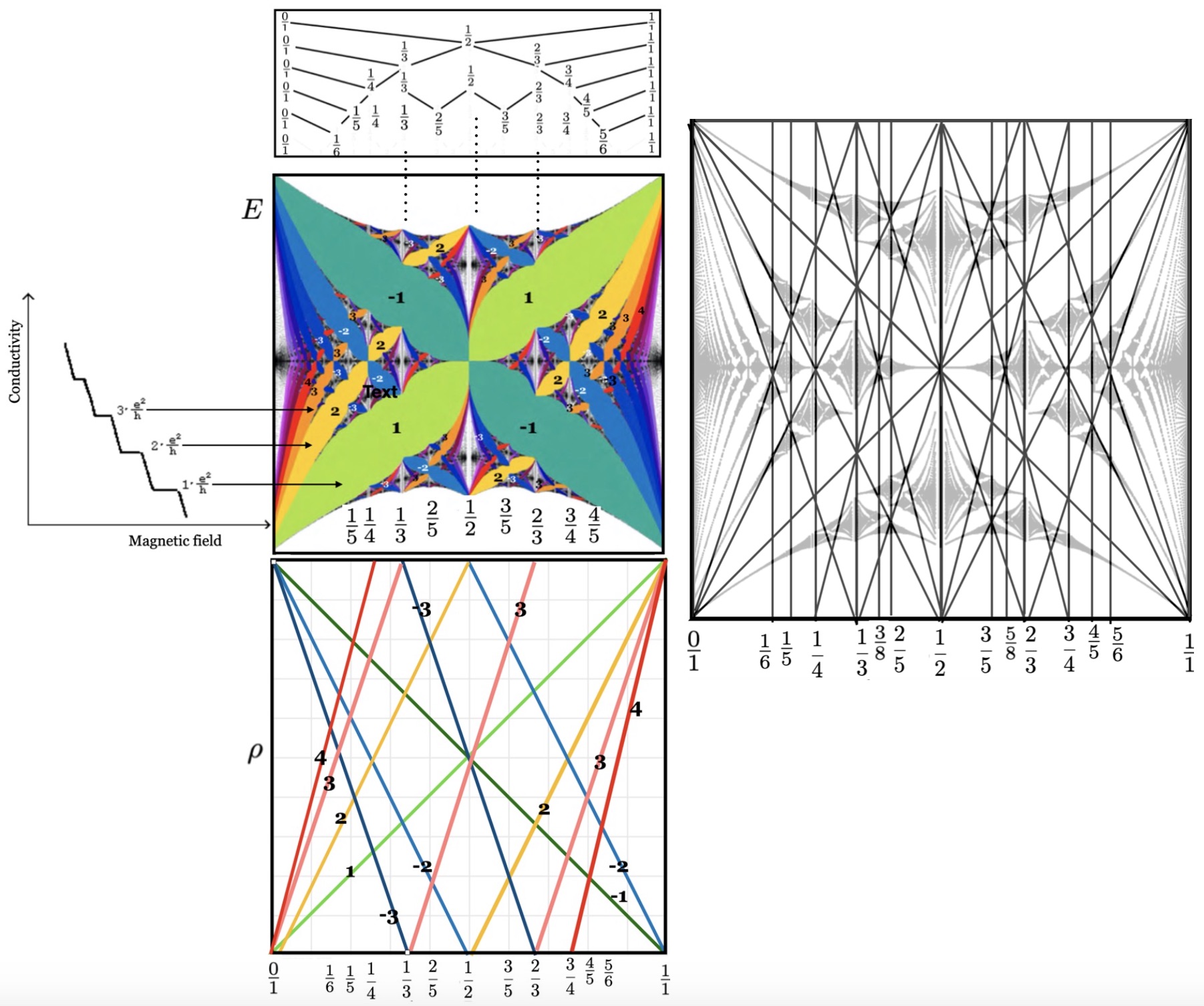} 
\leavevmode \caption{ Sketch of plateau of Hall conductivity (left) and the Butterfly graph where gaps are labeled with quantum numbers of the Hall conductivity as shown by the arrows.  The three vertical panels at the center portray the  relationship between the butterfly graph ( central panel)  and abstract mathematical  hierarchical sets- namely the Farey tree ( top panel) and the Wannier diagram ( the bottom panel). In the rightmost panel, the butterfly graph  and the Wannier diagram are superimposed.  This illustrates how the Wannier diagram provides a geometrical representation of the butterfly graph as the boundaries and the centers of the butterflies are encoded in the Wannier trajectories.}
\label{F}
\end{figure}

 In this paper, we further examine the role of the number theory in this quantum system where competition between crystalline lattice and cyclotron radius  lies at the very heart of the emergent hierarchical spectrum. 
We show that 
the Farey tree - a hierarchical set that generates all prime fractions between zero and one and the Wannier diagram which provides a simple representation of the butterfly graph are closely related. Figure ({\ref F}) highlights the number theoretical aspect of the  butterfly graph - a fractal made up of integers where the integers represent the quantum numbers of Hall conductivity.  They appear as the slopes of straight line  trajectories in a unit square, the Wannier diagram. Furthermore,  the nesting of the  butterfly spectrum is encoded in the Farey tree hierarchy.  
 Relationship between the Farey tree and the Wannier diagram  is shown to result in a hierarchical lattice of trapezoids - dubbed `` Farey-Wannier lattice". Every butterfly in the butterfly graph
can be paired with a trapezoid in the  lattice, thus encoding all the number theoretical characters of the butterfly. The lattice excludes certain trapezoidal configurations that do not represent butterfly patterns. This feature  falls outside number theoretical framework  is found to be described by a simple rule where the forbidden configurations correspond to  ``higher order" violation of the symmetry of the butterflies. In addition to Harper model\cite{Harper} - the nearest-neighbor
(NN) tight binding model of two-dimensional Bloch electrons in  magnetic field, we also discuss a generalized next-nearest-neighbor (NNN) model\cite{NNN, NNNT, NNN1,NNN2, Math2}   and show that Farey hierarchy prevails in characterizing the hierarchical  structure of the energy spectrum. In the latter case,  the Farey-Wannier lattice regroups to form hexagonal cells that describe butterflies with somewhat different number theoretical characteristics. In our limited exploration, the butterfly recursions for the generalized Harper equation is found to be  described by the  scaling factors $\zeta= [n^*+1; \overline{ 1, n^*}]$ that underlies the Harper equation. However, the renormalization scheme supports the possibility of new universality classes.\\

In section II, we began with a brief review of the butterfly Hamiltonian and its relation with the Wannier diagram and  the Farey tree.   Section III shows how the Farey tree and the Wannier diagram are related
and discusses Farey-Wannier  hierarchical lattice of trapezoids that provides new insight towards its relation to the butterfly graph.  As described in section IV, not all Wannier trajectories lead to the formation of butterflies.
Section V discusses generalized Harper model with new species of butterflies. In Appendix, we  show that self-similar Farey hierarchy can be described as a conformal map -
a M\"{o}bius transformation that encodes the  self-similar butterfly hierarchies for both the NN and the NNN model. 

\section{The Butterfly Graph}

Butterfly Hamiltonian, the Harper model, is a simple model of two dimensional non-interacting, spin-less electrons in a perpendicular magnetic field  $B$ where electrons moving in a square lattice can hop only to its nearest neighbor sites.
The key parameter in the problem is the magnetic flux per unit cell of the lattice in the units of the flux quanta
$\phi=  \frac{B a^2}{\hbar/e}$. 
In its simplest form, the model can  be written as\cite{MW1} as,
 
 \begin{equation}
 H =  \cos x +  \cos p, \,\,\,\, \ [x, p] = i \phi,
 \end{equation}
 
That is, butterfly graph lives in space of energy E and the effective Planck's constant $\phi$. The graph resembled a butterfly with a highly intricate recursive structure ,
 consisting of nothing but copies of itself - the ``sub-butterflies", nested infinitely deeply.\\
 
  For a rational  flux $\phi=\frac{p}{q}$,  the butterfly spectrum consists of  $q$ bands,  separated by $(q-1)$  gaps that form the wings of the butterfly.  For $q$ even, the two bands touch at the center of the spectrum, that is at $E=0$. For irrational case, the spectrum is a Cantor set where the allowed values of  the energy is set of zero measure. This is known as the `` Ten Martini Problem" --
the name was coined by Barry Simon in this 1982 article\cite{Ten}, originated from the fact that Mark Kac has offered ten martinis to anyone who solves it.\\
  
When $\phi=0$  or when $\phi = 1$, the energy spectrum of the Harper equation is a single interval which happens to be $[-4,4]$. This indicates that we can define ``sub-images" by identifying one edge with a single band of the spectrum, bounded by a gap on either side. For example, we may take one particular band of the spectrum when $\phi_L= \frac{p_L}{q_L}$ as forming the left-hand edge of the sub-image. As we increase $\phi$ away from its initial value of $\phi_L$, the spectrum becomes very complex, but the gaps which separate the sub-image from the rest of the spectrum persist. We may find that when $\phi$
 reaches another rational value, $\phi_R = \frac{p_R}{q_R}$, the complex spectrum reforms into a single band, forming a sub-butterfly. 
 In the Harper model, the structure of the sub-spectrum in the region between $\phi_L$ and $\phi_R$ is well- approximated by a distorted version of the original Hofstadter butterfly\cite{MW1, SW}. As we describe later,
 in a generalized NNN-Harper model, new species of butterflies appear which are not related to the main butterfly. However,  both the NN and NNN models are described by the same renormalization equations
 as both are embedded in the Farey tree.
 
\subsection{ Wannier Diagram - Butterfly Skeleton }

The Wannier diagram, named after Gregory Wannier   who in $1978$ revisited\cite{W,CW} the problem of a crystal in a magnetic field shortly after the discovery of the butterfly graph. It  provides a simple representation of the spectrum by  labeling all the gaps of the spectrum with two integers $(\sigma, \tau)$, expressed as a linear Diophantine equation,

\begin{equation}
r = p \sigma + \tau q, \,\,\  \rho \equiv \frac{r}{q} = \sigma \phi + \tau .
\label{Dio}
\end{equation}

Here $r$ labels the $r^{th}$ gap of the spectrum for a rational magnetic flux $\phi = \frac{p}{q}$ and $\rho$ is the density of the electrons or the fraction of total number of states below Fermi energy.
The $\rho$ vs $\phi$ plot can be viewed as representing   ``butterfly skeleton" as various ``Wannier trajectories"  representing the gaps of the butterfly, shrink to straight lines.
For a given set of values for $(r, p, q)$,  there are infinitely many solutions to any such Diophantine equation. Indeed, it is easy to see that if $(\sigma, \tau$) is a solution of equation  then so is  $( \sigma + n q, \tau-np)$, $ n= 0, \pm 1, \pm 2.....$.It turns out that for the rectangular lattice, what we want is the smallest possible $\sigma$ (in absolute value).\\
 
In  $1982$,  the linear Diophantine equation got a big boost after  Thouless et al\cite{TKNN} showed that the integer $\sigma$ in the equation (\ref{Dio}) represents quantum number of Hall conductivity and  has topological origin. Following this important discovery for which David Thouless was awarded Nobel prize in (2016), 
Eq.  (\ref{Dio}) has been subject of various studies\cite{T83, Mac, JM, DS, Dana} and is also referred as the 
  the ``gap labeling theorem\cite{Simon}. In recent experimental investigation of the butterfly spectrum\cite{Dean},  the calculation of $(\sigma,\tau)$ from the measurement of
 the filling-fraction $\rho$, emerged as the key factor in providing laboratory glimpses of butterfly fractal.\\

 \subsection{ Farey Tree and the Butterfly Graph}
 
 Although the connection between the butterfly graph and the Wannier diagram has been known since $1978$, the relationship between the hierarchical nature of the butterfly graph and the Farey tree was first pointed out in $2016$\cite{book,Sat16}. In $2020$, these  empirical results were derived\cite{SW}  using previously known renormalization group\cite{MW1}, thus establishing the fact that the quantum mechanics of the Bloch electrons in a magnetic field is intertwined with various number theoretical results. Below we briefly review the Farey tree construction and its relationship with the butterfly graph.
 
 \begin{figure}[htbp] 
\includegraphics[width = .3\linewidth,height=.35 \linewidth]{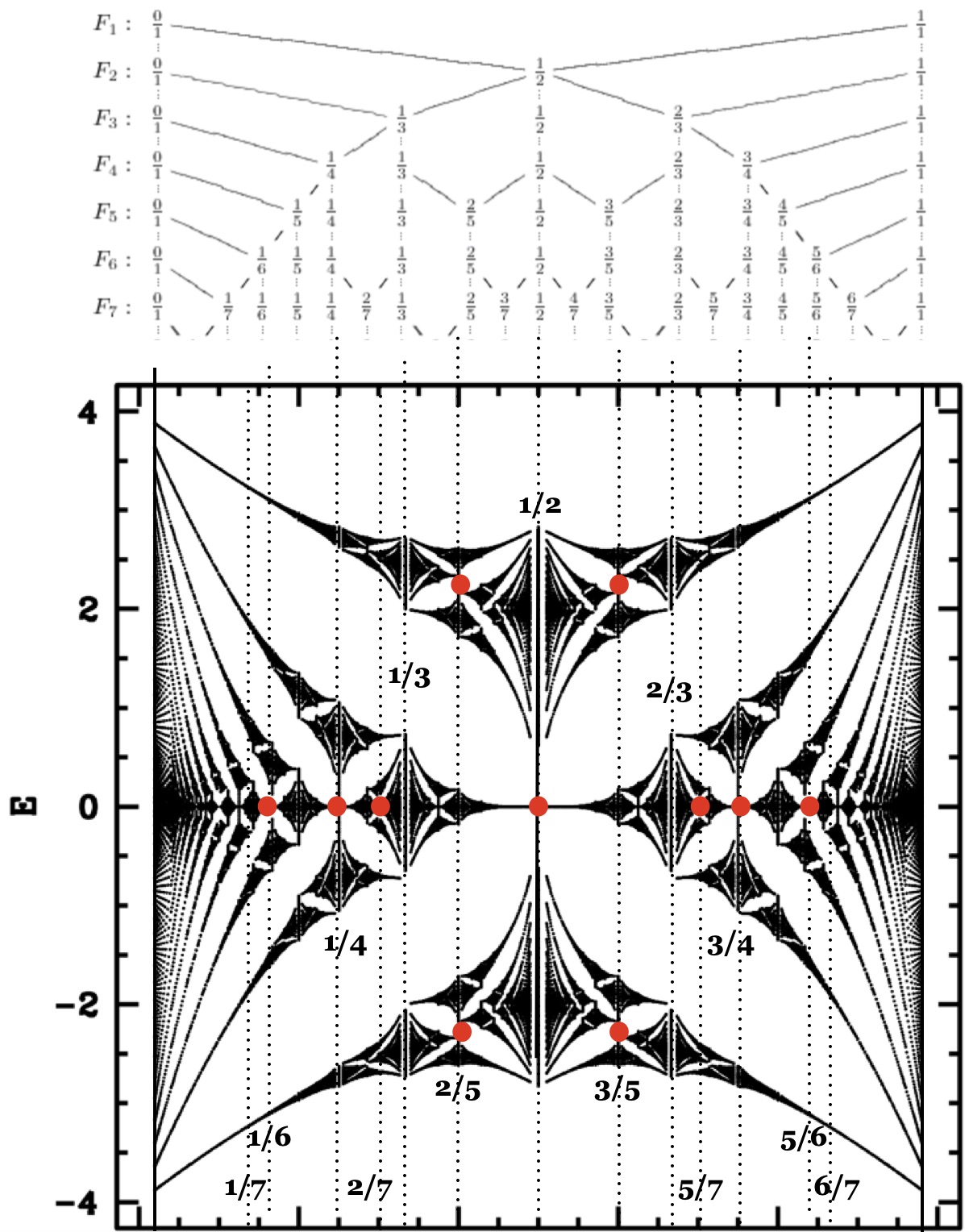}
\leavevmode \caption{ The Farey Tree  and the butterfly graph.The left-right boundaries and the center of every sub-butterfly in the graph can be labeled with a friendly Farey triplet.}
\label{FTB}
\end{figure}

Discovered by Adolf Hurwitz in $1894$,  Farey tree  generates all primitive rationals between $0$ and $1$. As shown in Fig. (\ref{FTB}), this hierarchical tree-like structure  builds the entire set of rationals by starting with $0$ and $1$. 
Given any two fractions $\frac{p_L}{q_L}$ and $\frac{p_R}{q_R}$ that satisfy 
\begin{equation}
p_L q_R - p_R q_L = \pm 1,
\label{Flr}
\end{equation}

then $p_L $ and $q_L$ are coprimes and so is $p_R$ and $q_R$. This is because any common factor of $p_L$ and $q_L$ must divide the products $p_Lq_R$ and $p_R q_L$ and hence the difference $p_L q_R - p_R q_L = \pm 1$.  Any two fractions satisfying  Eq. (\ref{Flr}) are two neighboring fractions in the Farey tree and are known as the {\it friendly fractions}.  Farey tree is constructed by applying
 the `` Farey sum rule" to  $\frac{p_L}{q_L}$ and $\frac{p_R}{q_R}$- the Farey parents that gives a new  fraction $\frac{p_c}{q_c}$ - the Farey child:
\begin{equation}
\frac{p_c}{q_c} = \frac{p_L+q_R}{q_L+q_R}.
\label{Fc}
\end{equation}
Analogous to the friendly pair $\frac{p_L}{q_L}$ and $\frac{p_R}{q_R}$, $\frac{p_c}{q_c}$ also forms friendly pair with each of its parents $\frac{p_L}{q_L}$ and $\frac{p_R}{q_R}$, satisfying the following two equations,
\begin{eqnarray}
 p_L q_c - p_c q_L  & =  & \pm 1\\
 \label{Fcl}
 p_c q_R - p_R q_c  & = & \pm 1
 \label{Fcr}
\end{eqnarray}

This implies that $p_c$ and $q_c$ are also coprime. In other words, the entire Farey tree consists of all  fractions  $\frac{p}{q}$ where $p$ and $q$ are coprime.
These equations  define a
Farey triplet denoted as $[\frac{p_L}{q_L}, \frac{p_c}{q_c}, \frac{p_R}{q_R}]$ which will be referred   as the ``{\it friendly Farey triplet}".\\ 

Importance of friendly Farey triplets  in butterfly spectrum was pointed out in our recent studies\cite{book, SW, Sat21}. It was  shown that
the magnetic flux values corresponding to  friendly triplets $[\frac{p_L}{q_L}, \frac{p_c}{q_c}, \frac{p_R}{q_R} ]$ in the Farey tree form the left  boundary, the center and the right flux boundaries of the butterflies,
encoding the hierarchical structure of the butterfly graph as shown in Fig. (\ref{FTB}). 
In other words, butterfly graph is an incarnation of the Farey tree adorned with butterflies.

 \section{ Relating Farey Tree and the  Wannier Diagram }
 
\begin{figure}[htbp] 
\includegraphics[width = 0.8 \linewidth,height=.5\linewidth]{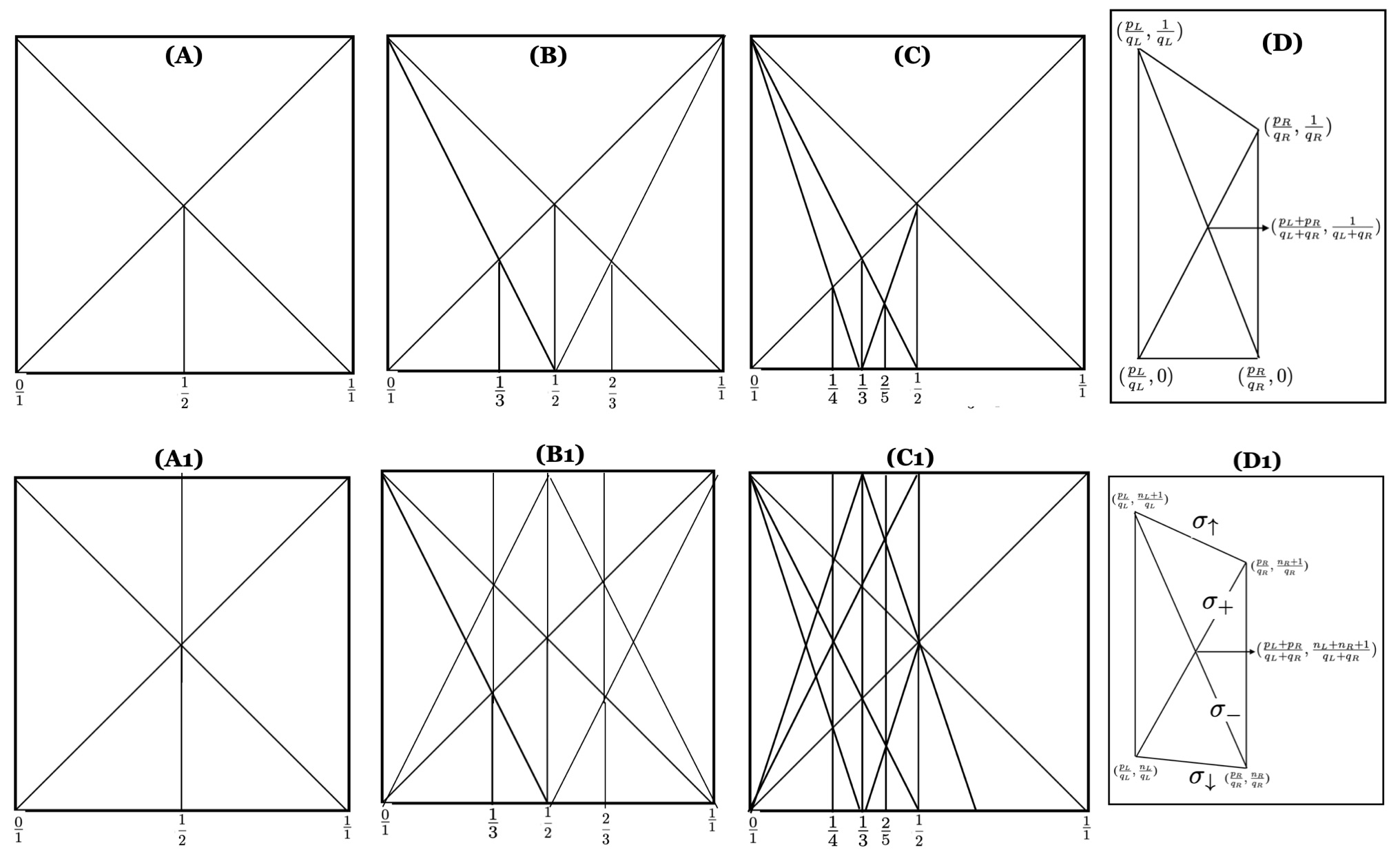}
\leavevmode \caption{ A-C show an alternative way to construct the Farey tree  using straight lines in a square. Panel D shows a general trapezoid, with the coordinates of the four corners are labeled.  Lower panel: $A_1-D_1$ show the corresponding  Wannier diagram.}
\label{FWall}
\end{figure}

We now describe an alternate way  to construct the Farey tree and show that this geometrical construction is intimately related to the Wannier diagram. Fig. (\ref{FWall}) shows the construction of this diagram in stages, which we summarize below.
\begin{itemize} 
\item (1) Start with drawing a unit square  and its diagonals. 
\item (2) Draw a vertical line from the intersection point of the diagonals down to the bottom edge of the square.  Starting with two rational numbers $\frac{0}{1}$ and $\frac{1}{1}$,
 It gives a new rational number $\frac{1}{2}$.  The process generates two trapezoids:
one to the left of $\frac{1}{2}$
with parallel lines at $\frac{0}{1}$ and $\frac{1}{2}$ and another to the right of $\frac{1}{2}$  with parallel lines at $\frac{1}{2}$ and $\frac{1}{1}$.
\item (3)Repeat the above process with each trapezoid: that is draw its diagonals and then draw  vertical lines to the bottom from the intersection point of the diagonals of the trapezoids. This gives the Farey fraction $\frac{1}{3}$ and $\frac{2}{3}$.  Vertical line from each of the fraction generates two new trapezoids, one to the left  and the other to the right of that fraction. 
\item (4) Continue this process: with each new trapezoid, draw its diagonals and the vertical line from the point of intersection of the diagonals.
 This will generate all rationals because the vertical lines from the diagonals of the trapezoid  formed by two parallel lines at $\frac{p_L}{q_L}$ and $\frac{p_R}{q_R}$ meet the bottom edge at $\frac{p_L+p_R}{q_L+q_R}$ - the Farey sum of the two fractions. This is  shown in panel (D).
\item (5) In general, for every trapezoid so formed with two parallel lines at  friendly fractions $\frac{p_L}{q_L}$ and $\frac{p_R}{q_R}$, the y-coordinates of the upper left and the upper right corners of the trapezoid are $\frac{1}{q_L}$ and $\frac{1}{q_L}$. The coordinate of the intersection of the diagonals is $( \frac{p_L+p_R}{q_L+q_R}, \frac{1}{q_L+q_R})$. By induction, this proves that the length of every vertical  line of the trapezoid  at
fraction $\frac{p}{q}$ is $\frac{1}{q}$. 
\end{itemize}
From this geometrical construction of the Farey tree, it turns out that the Farey tree is closely related to the Wannier diagram.  As illustrated in  the  upper panels of figure (\ref{FWall}), it involves symmetrization of the Farey construction about  the $y=\frac{1}{2}$ line of the square.
In other words, as  the unit square is transformed into a two-torus, the resulting geometrical figure is the Wannier diagram where the $x-y$-axes are identified as the variables $\phi-\rho$ of the Wannier diagram.\\

In summary, starting with the geometrical construction of the Farey tree  (Fig. (\ref{FWall}) ),  Wannier diagram emerges in two steps. Firstly,  all vertical lines
are extended up to the upper edge of the unit square,  which is identified with the $\rho=1$ line of the Wannier diagram. Secondly, new lines are added so that the entire configuration is symmetrical about $\rho=1/2$. In other words the density  $\rho$ as a function of $\phi$ satisfies the 
condition $\rho(\phi)= \rho(1-\phi)$. \\

The key point to be noted is that all the slanting lines  in Fig. (\ref{FWall}) have integer slopes and integer intercepts when the parallel lines of the trapezoid are at friendly fractions and height of each parallel line at fraction $\frac{p}{q}$ is $\frac{1}{q}$. To see this, consider a general trapezoid , shown in panel (D1) of Fig. (\ref{FWall}).  For example, 
the diagonal line with positive slope, denoted as  $\sigma_+ = \pm ( \frac{n_R+1}{q_R}-\frac{n_L}{q_L})/(\frac{p_R}{q_R}-\frac{p_L}{q_L}) = (n_R+1) q_L-n_L q_R$ as $p_L q_R -p_R q_L = \pm 1$. Table (\ref{T1}) lists slopes and intercepts of all non-parallel lines of the trapezoid.
 \begin{table}
\begin{tabular}{| c | c |  c  | c | }
\hline
$\sigma_+,\,\ \tau_+$ \,\,  & $\pm [ n_L q_R - (n_R+1) q_L] ,\,\ \mp [n_L p_R - (n_R+1) p_L]$    \\ 
\\
\hline
$\sigma_-,\,\ \tau_-$ \,\,  & $\pm [ (n_L+1)  q_R - n_R q_L] ,\,\ \mp [(n_L+1) p_R - n_R p_L]$    \\ 
\\
\hline
$\sigma_{\uparrow},\,\ \tau_{\uparrow}$ \,\,  & $\pm [ (n_L+1) q_R - (n_R+1) q_L] ,\,\ \mp [(n_L+1) p_R - (n_R+1) p_L]$    \\ 
\\
\hline
$\sigma_{\downarrow},\,\ \tau_{\downarrow}$ \,\,  & $\pm [ n_L q_R - n_R q_L] ,\,\ \mp [n_L p_R - n_R p_L]$    \\ 
\\
\hline
\end{tabular}
\caption{  The slopes and the y-intercepts   of the  diagonals denoted as  $(\sigma_{\pm}, \tau_{\pm})$ and non-parallel  lines $(\sigma_{\uparrow, \downarrow}, \tau_{\uparrow,\downarrow})$ of the trapezoidal cells shown in panel $D_1$ of Fig. (\ref{FWall}) where $p_L q_R - p_R q_L = \pm 1$. }.
\label{T1}
\end{table}

\begin{figure}[htbp] 
\includegraphics[width = 0.4 \linewidth,height=.42\linewidth]{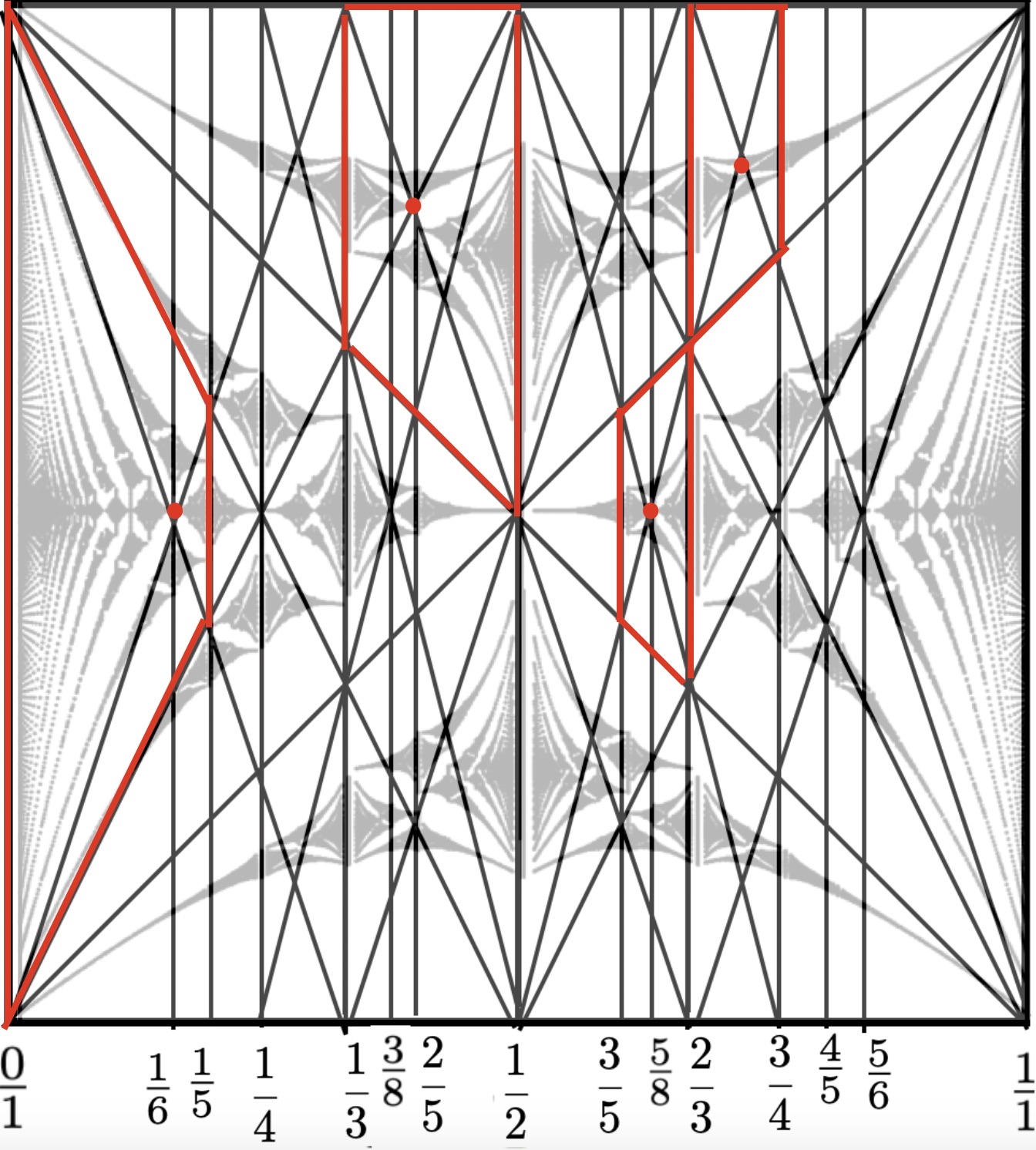}
\leavevmode \caption{ Butterfly graph with four of its sub-butterflies identified by (red) dots at the center and the corresponding trapezoidal cells  ( in red) of the ``Farey -Wannier lattice" }
\label{FWL}
 \end{figure}

\section {Farey-Wannier Lattice and the Butterfly Graph }
The Wannier trajectories form  a very special type of a hierarchical lattice made up of trapezoidal cells in every Farey interval  $[\frac{p_L}{q_L}-\frac{p_R}{q_R}]$ where the Farey fractions  $\frac{p_L}{q_L}$ and $\frac{p_R}{q_R}$ are neighbors in the Farey tree, ie. they satisfies the friendly fraction condition $ p_L q_R - p_R q_L = \pm 1$.  The points of intersections of the diagonals of the trapezoids represent the center of the butterfly. In this lattice,  all slanting lines have integer slopes and also integer intercepts. As described below in section  IV-B, not all trapezoidal cells correspond to butterflies. Such a hierarchical lattice where every trapezoid represents a butterfly will be  dubbed  as the ``Farey-Wannier lattice" as shown in  Fig. (\ref{FWL}) where the butterfly graph is superimposed on the Wannier diagram.\\
\\
\\
\\
\subsection{ Nests and Chains}

In general,  a friendly interval $[\frac{p_L}{q_L}-\frac{p_R}{q_R}]$ where $(q_L < q_R)$,  consists of a stacks  $q_L$ trapezoids and $(q_R-q_L)$ triangular regimes.
As we look at the hierarchical lattice with higher order Farey fractions, the triangular regimes get packed with an infinite chain of trapezoids of different widths that asymptotically approaches zero. 
   Figures ~(\ref{BW1}) shows a slab of hierarchical lattice in a friendly  interval $[1/2-1/3]$ with trapezoids and triangles which overlays the  corresponding butterfly graph.  This interval consists of two trapezoids 
   representing two  butterflies. A triangular region sandwiched between  the two trapezoids represent a chain of butterflies.  With higher order Farey fractions, each butterfly gets infinitely nested.
   This is further illustrated in Figure ~(\ref{BW3}) with three friendly intervals $[2/7-1/3], [1/3-2/5], [2/5-3/7]$.  As described below, not all trapezoids represent butterflies.
   
   \subsection{Minimal Symmetry violation}
 
\begin{figure}[htbp] 
\includegraphics[width = 0.6 \linewidth,height=.3 \linewidth]{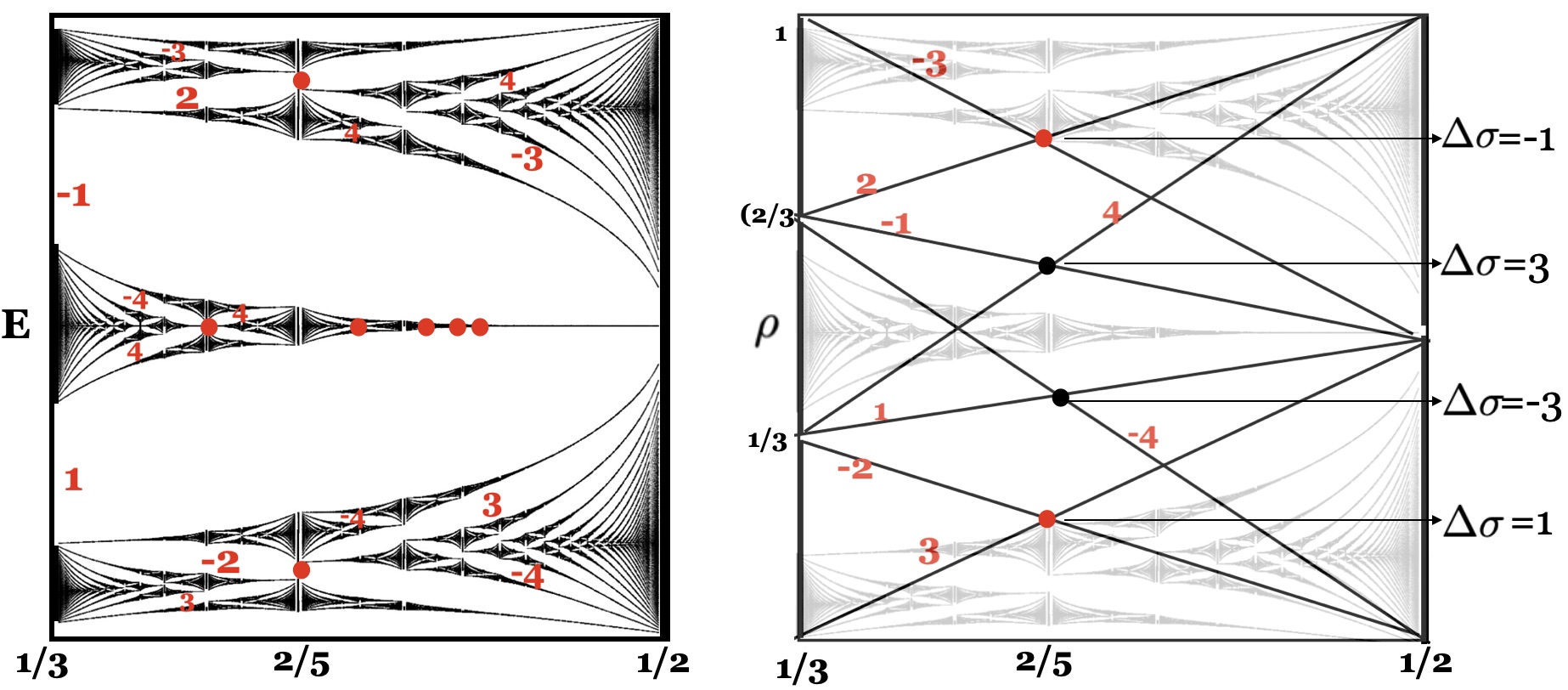} 
\leavevmode \caption{  In the friendly interval $[1/3,1/2]$, left panel shows a chain of butterflies at the center and a butterfly at the upper and the lower edges. The right panel shows the corresponding Wannier diagram, superimposed on the butterfly graph in that interval. Figure  illustrates how quantum mechanics {\it improvises}  on the number theory encoded in the Wannier diagram  to create butterflies as
it chooses only certain trapezoids (shown with red dots) that represent butterflies. Trapezoids whose centers are shown in black dots are rejected by the quantum mechanics determining the butterfly graph. The flux interval $[1/3,1/2]$  can support only two butterflies and they exhibit  horizontal mirror symmetry about $\rho=\frac{1}{2}$,  corresponding to the asymmetry parameter $\Delta \sigma = 1$ and $-1$ respectively.  Wannier trajectories that are shown to intersect at black dots  correspond to $\Delta \sigma = \pm 3$ do not correspond to any butterflies. In other words, the butterflies exhibit minimum violation of mirror symmetry.}
\label{BW1}
\end{figure}

\begin{figure}[htbp] 
\includegraphics[width = 0.75\linewidth,height=.35 \linewidth]{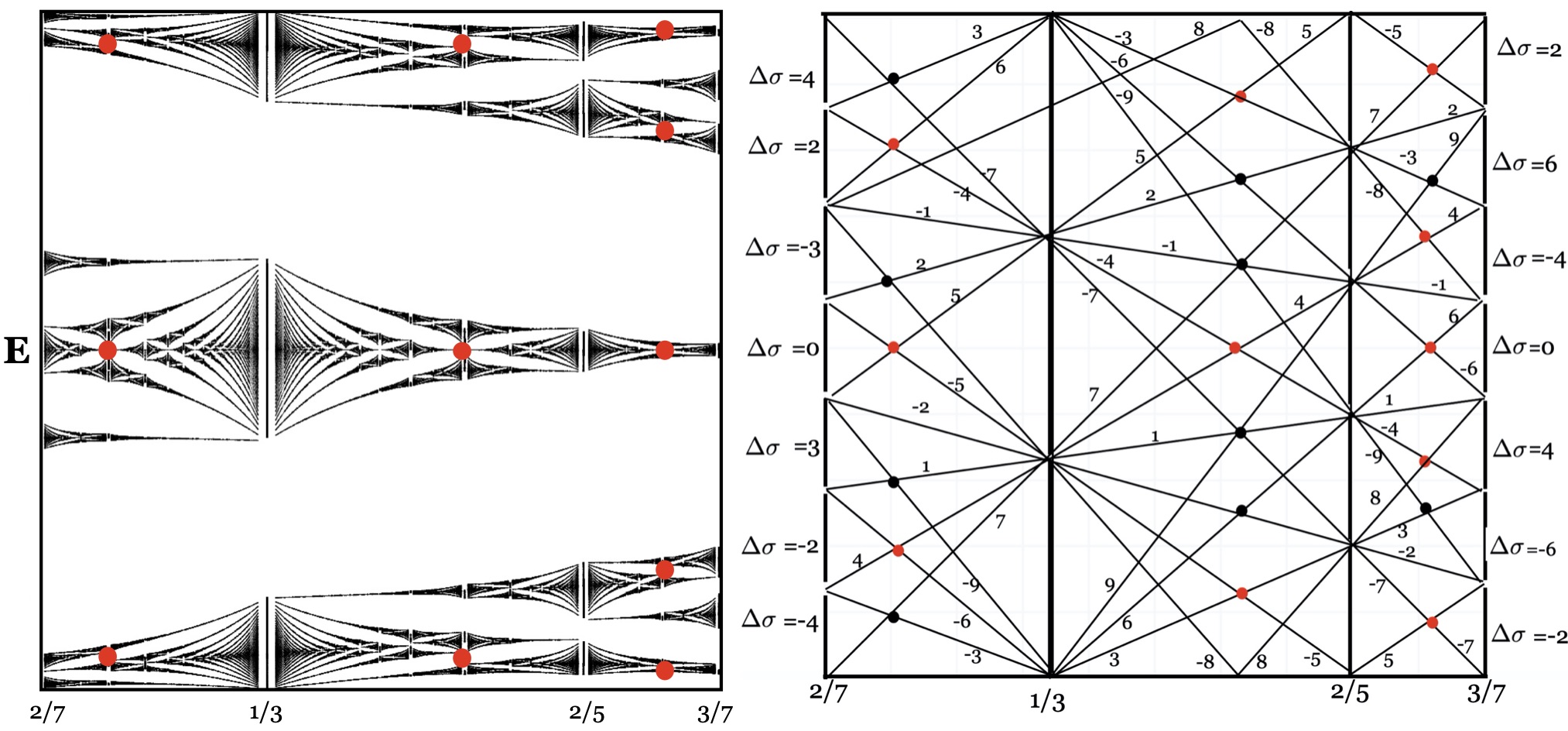}
\leavevmode \caption{ Butterfly and the corresponding Wannier diagram in three friendly magnetic flux intervals $[2/7,1/3], [1/3, 2/5]$ and  $[2/5,3/7]$.  Each trapezoid is uniquely determined by a dot ( red or black) at the intersection of the two diagonals. Red  dots represent  the configurations that correspond to the butterflies and can be paired with the red dots in the butterfly graph on the left.  Configurations corresponding to black dots do not appear - a feature determined not by the number theoretical arguments. Trapezoids with black dots are  forbidden configurations that  correspond to higher values of $\Delta \sigma$ - the asymmetry parameter, shown explicitly for the left and the right  trapezoids. For central interval $[1/3, 2/5]$, the values of $\Delta \sigma= 2, -4, 6, 0, 4, -6$ ( from top to bottom) . The trapezoids with horizontal mirror symmetry have $\Delta \sigma =0$. }
\label{BW3}
\end{figure}

   Figures ~(\ref{BW1}) and  ~(\ref{BW3})  illustrates an arbitrariness in the choice of selecting trapezoidal and triangular regimes, as we seek one to one correspondence between  the trapezoids and the butterflies.
 In constructing a Farey-Wannier lattice representing an isomorphism between the hierarchy of trapezoids and the butterflies, we now  address the key question of  what determines  the {\it right} choice of grouping trapezoidal and triangular regimes of the lattice, shown with red dots in the figure. The configurations corresponding to black dots are not {\it used} in the butterfly graph and are rejected.
A close inspection of the correspondence between the Farey-Wannier lattice and the butterfly graph shows that the trapezoids cells that do not represent a butterfly can be singled out by a parameter that characterizes the {\it degree of violation of horizontal mirror symmetry}.  Such a symmetry corresponds to the difference in the magnitude of Chern numbers $( \sigma_+, \sigma_-)$ as for the central butterflies that exhibit mirror symmetry $|\sigma_+|=|\sigma_-|$. We define a parameter $\Delta \sigma$ as:
  
 \begin{equation}
 \Delta \sigma = |\sigma_+| - | \sigma_-| =  |(2n_R+1)  q_L-(2n_L+1)  q_R|
 \end{equation}
 
 For central trapezoids , $\Delta \sigma=0$  as $n_R = \frac{q_R-1}{2}, n_L = \frac{q_L-1}{2}$.
 In a given interval $[\frac{p_L}{q_L}-\frac{p_R}{q_R}]$, defined by the friendly fractions,  there exists $q_L$ butterflies, ( when $q_L < q_R$ ) each characterized by a unique $\Delta \sigma$.
Therefore, $\Delta \sigma$ can be taken as a measure of  the asymmetry of the trapezoid ( and the corresponding butterfly) as higher the value of $\Delta \sigma$, greater is the degree of violation of horizontal mirror symmetry. As illustrated in  figures, the
 trapezoids that are not paired with butterflies correspond to higher values of $\Delta \sigma$. In other words, given all possible trapezoids in a given rectangular strip, each labeled with a unique value of $\Delta \sigma$, nature {\it uses} trapezoids with smallest possible value to represent butterflies in the butterfly graph.
 \\
 \\
 \\
 \\
 \\
 \\
 \\
 \\
 \\
 \\

 \section{ New Species of Butterflies}
 
We now discuss the number theoretical properties of the energy spectrum in a generalized Harper Model\cite{NNN, NNNT, NNN1, NNN2}  described by the tight binding Hamiltonian, 
   \begin{equation}
 H =  t_a \cos p +  t_b \cos x +  t_{ab} [ \cos(x-p) + \cos(p-x) ]
 \end{equation}
 
 Here $t_a$ and $t_b$ are NN hopping along the $x$ and the $y$ direction and $t_{ab}$ defines the NNN hopping between the diagonals of the square lattice. \\

 As we tune the parameters, the energy spectrum shows changes, although the  patterns resembling butterflies persist. 
 We examine the role of number theory  in characterizing the energy spectrum with a  key question  whether  the Farey sum rule ( Eq. (\ref{Fc}) ) continues to define the butterfly-like spectrum.\\
 
  In our  study of the NNN model spectrum,  the Farey hierarchy was found to prevail. However, in addition to butterflies that obey Farey sum rule ,  which we refer as type-I butterflies,
  there are new types of butterflies- type-II and type-III butterflies,
   where  the Farey triplet $[\frac{p_L}{q_L}, \frac{p_c}{q_c}, \frac{p_R}{q_R}]$ does not form a friendly triplet. The modified Farey sum rules for these new species of butterflies are given below.
  \\
  \\
  For Type-II butterflies,
 \begin{eqnarray}
 p_L q_R -p_R q_L = \pm 2,\, \,\ p_c q_R -p_R q_c = \pm 1,\, \,\ p_c q_L -p_L q_c = \pm 1,
  \label{FS1}
  \end{eqnarray}
  Consequently,  $p_c =\frac{p_L+p_R}{2}$ and $q_c =\frac{q_L+q_R}{2}$.\\
  
  For Type-III butterflies:
   \begin{eqnarray}
 p_c q_R -p_R q_c = \pm 2 ,\, \,\ p_c q_R -p_R q_c = \pm 1,\, \,\ p_L q_R -p_R q_L = \pm 1.
  \label{FS2}
  \end{eqnarray}
  
  Therefore, $p_L  = \frac{p_c-p_R}{2}$ and $q_L= \frac{q_c-q_R}{2}$\\
  
  or 
   \begin{eqnarray}
  p_c q_L -p_L q_c = \pm 2 ,\, \,\ p_c q_L -p_L q_c = \pm 1,\, \,\ p_L q_R -p_R q_L = \pm 1,
  \label{FS3}
  \end{eqnarray}
  
  and therefore $p_R  = \frac{p_c-p_L}{2}$ and $q_R= \frac{q_c-q_L}{2}$\\
  
   Therefore, $ p_x q_y -p_y q_x = \pm 1$  is true for two of the three pairs from $(L, c, R)$. For the third pair,
 $ p_x q_y -p_y q_x = \pm 2 \equiv D$. That is, among the three pairs of
magnetic flux fractions at the left and right boundaries and at the center,  two pairs are NN in Farey tree and one pair that is NNN in the tree.\\ 
\\
\\
\\
\\
\\
\\

 \begin{figure}[htbp] 
   \includegraphics[width = 0.85 \linewidth,height=.65\linewidth]{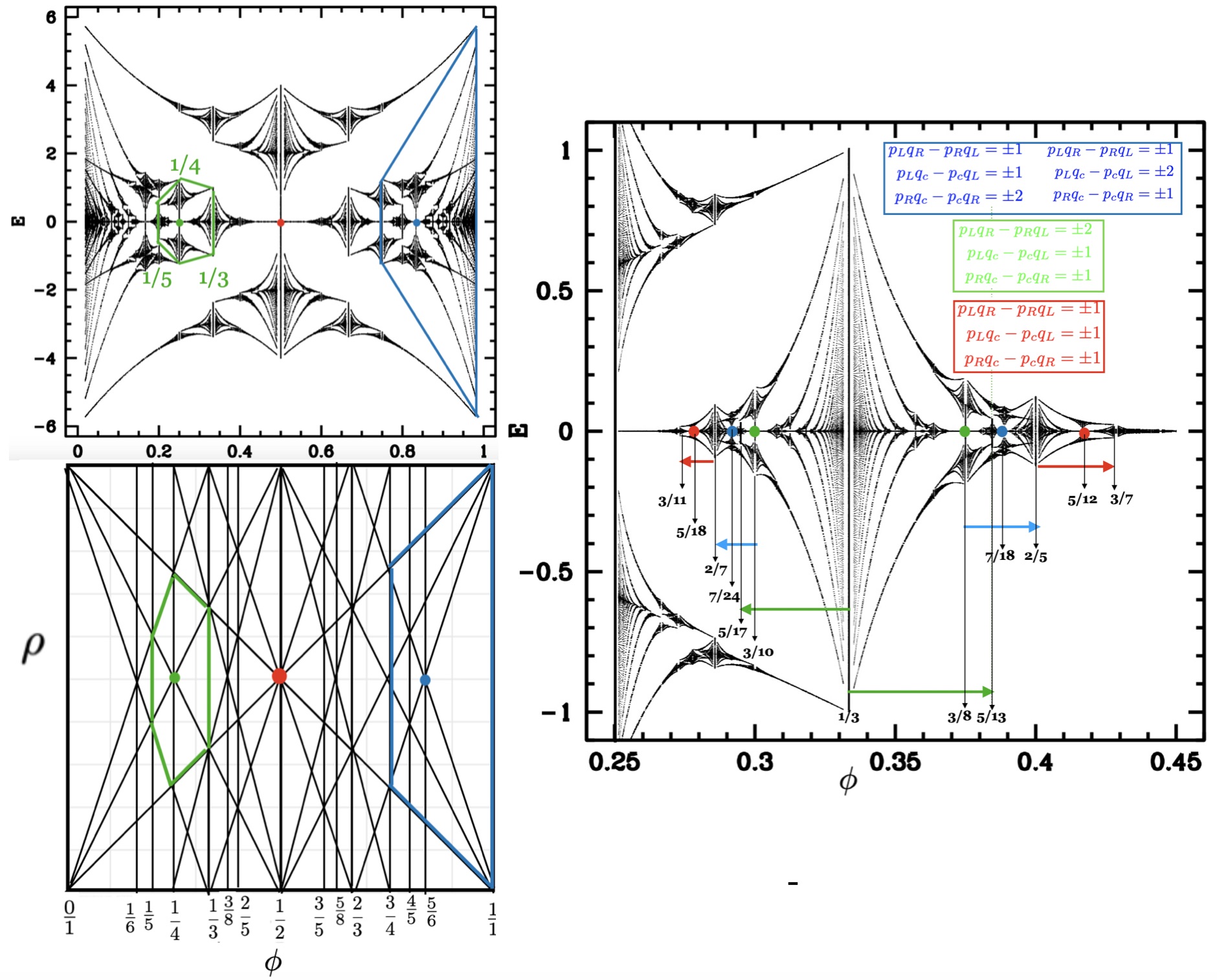}
\leavevmode \caption{  Upper left panel shows the butterfly graph for $t_{ab}=t_a=1$ and $t_b=0$. Examples of type-II  and type-III species are shown in blue and green.  The bottom left shows the corresponding Wannier diagram explicitly showing the  hexagonal and trapezoidal cells, color coded with the corresponding sub-butterflies in the upper panel.  Right panel shows a blowup of the central part of the butterfly showing the type-1 (red), the type-2 and type-3 ( blue and green) and the corresponding Farey relations, all color coded. The arrows, color coded, show the flux intervals for various butterflies.}
\label{NewB1}
\end{figure}

  Fig. (\ref{NewB1}) shows  an example of the spectrum for a special parameter values $ t_a = t_{ab}=1$ and $t_b = 0$ where butterflies satisfying modified Farey sum (\ref{FS1}) and (\ref{FS2}) are shown in green and blue. Distinction between the Type-I, Type-II  and Type-III, as illustrated in figure is further summarized in the Table (\ref{T2}).\\
   
  \begin{figure}[htbp] 
   \includegraphics[width = 0.6 \linewidth,height=.35\linewidth]{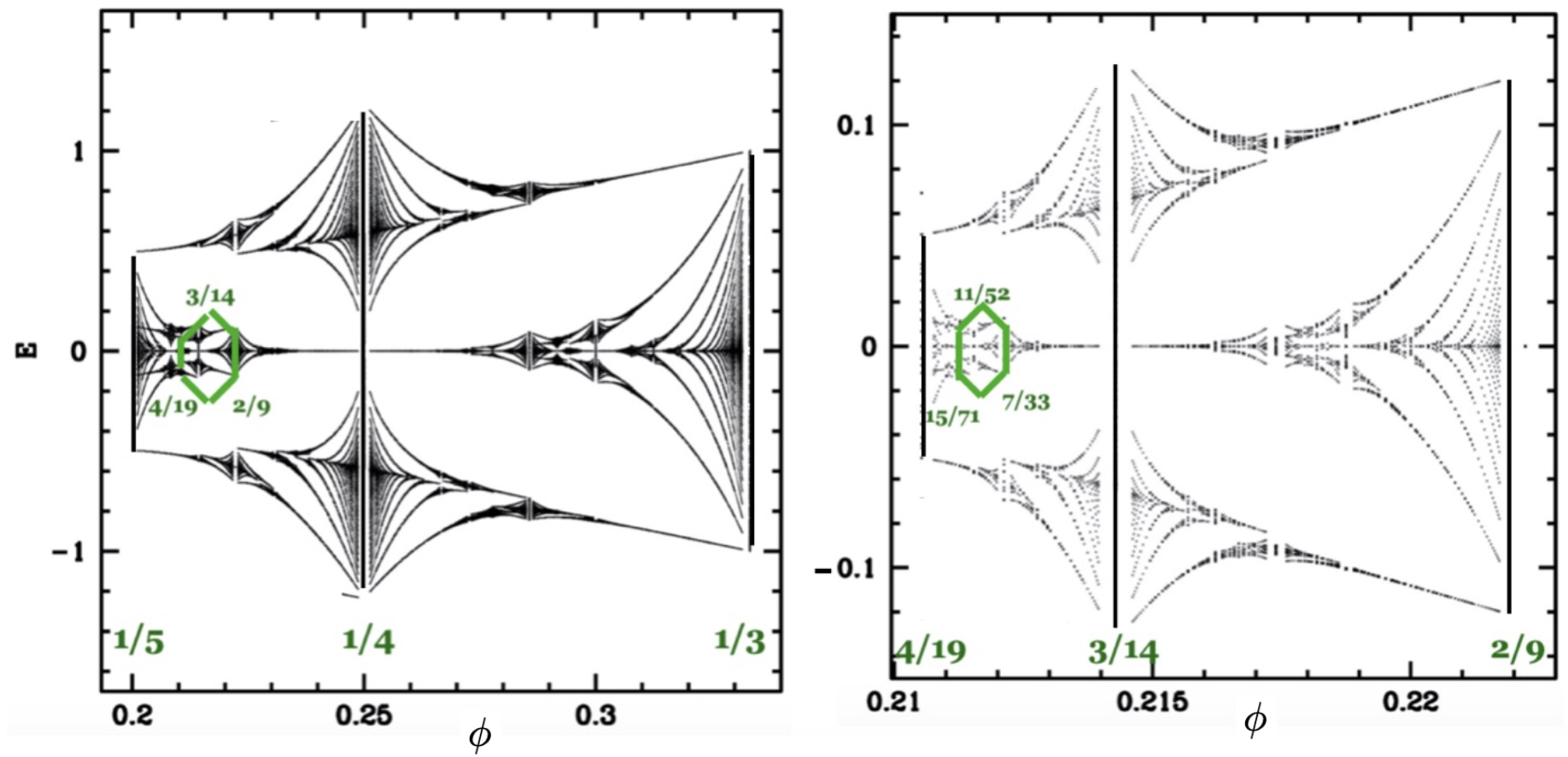}
\leavevmode \caption{  Two levels of blowups of the sub-butterfly ( upper-left green box )  in Fig. (\ref{NewB1}) residing at the center in flux interval $[1/3-1/5]$.}
\label{NewB2}
\end{figure}

 \begin{table}
\begin{tabular}{| c | c |  c  | c | }
\hline
 & 
\color{red}Type-I \,\, &  \color{green} Type-II   & \,\ \color{blue}Type-III \\ 
\hline
Farey Rule \,\,  & $ \frac{p_c}{q_c} = \frac{p_L+q_R}{q_L+q_R}$ \,\, & $ \frac{p_c}{q_c}  =   \frac{\frac{p_L+p_R}{2}}{\frac{q_L+q_R}{2}}$, &  $ \frac{p_L}{q_L}  =   \frac{\frac{p_c-p_R}{2}}{\frac{q_c-q_R}{2}}$,
or $ \frac{p_R}{q_R}  =   \frac{\frac{p_c-p_L}{2}}{\frac{q_c-q_L}{2}}$   \\ 
\hline
$(M_x,N_x)$  \,\,  & $M_L=M_R, N_L=N_R$ \,\, & $M_L \ne M_R, N_L \ne N_R$ & $M_L=M_R, N_L=N_R$  \\ 
\hline
Unit cell in Farey-Wannier Lattice \,\,  & Trapezoid & Hexagonal & Trapezoid  \\ 
\hline
Conformally Related to Main Butterfly \,\,  & Yes \,\, & No & No    \\ 
\hline
\end{tabular}
\caption{  Summary of type-I, type-II and type-III butterflies which are color coded in red, green and blue in the figure.}.
\label{T2}
\end{table}
 
\subsection{ Self-Similarity for the New  Butterflies} 

 Our previous studies have discussed in detail the self-similar hierarchies of type-I butterflies where magnetic flux interval for every sub-butterfly is related to the main butterfly by a M\"{o}bius transformation\cite{Sat21}.
 Appendix provides a broader perspective on the recursions that accommodates the type-I, the type-II and the type -III. Here we illustrate the self-similarity of the type-II butterflies with an example.

 Fig. (\ref{NewB2}) shows three levels of recursions for the the central band of the spectrum, corresponding to the  butterfly triplets $[1/3, 1/4,1/5] \rightarrow [4/19, 3/14, 2/9] \rightarrow [15/71, 11/52,7/33]$. 
 The renormalization equation can be constructed by relating the first two levels of the hierarchy, namely  $[1/3, 1/4,1/5] \rightarrow [4/19, 3/14, 2/9] $.  From the level-1 triplet $[1/3, 1/4,1/5] $, we pick any two fractions and relate it to the corresponding fractions in level-2 triplet  $[4/19, 3/14, 2/9]$, constructing a M\"{o}bius map as described in the Appendix. The resulting transformation is independent of the choice of the pair of friendly fractions used to construct the transformation and all three choices give the transformation matrix
 $T = \begin{bmatrix}  -1 & 1\\ -6 & 5 \end{bmatrix}$ as described in Eq. (\ref{t}). The matrix has eigenvalues $2\pm \sqrt{3}$ and hence
 scaling exponents fall with the same university class as type-I butterflies.
 
           We conclude with two important remarks about the the transformation $T$ that maps one pair of Farey fractions $(\frac{p_x}{q_x}, \frac{p_y}{q_y})$  to another pair 
          $(\frac{p^{\prime}_x}{q^{\prime}_x}, \frac{p^{\prime}_y}{q^{\prime}_y})$, preserving the determinant 
          $D= p_x q_y - p_y q_x=p^{\prime}_x q^{\prime}_y - p^{\prime}_y q^{\prime}_x$ and the order of fractions ,  namely $ \frac{p_x}{q_x} \rightarrow \frac{p^{\prime}_x}{q^{\prime}_x}$ and 
          $ \frac{p_y}{q_y} \rightarrow \frac{p^{\prime}_y}{q^{\prime}_y}$ .\\

 (1) The transformation matrix $T$  in Eq. (\ref{t}) describes the recursions for the Type-I, the  Type-II and the Type-III butterflies. Furthermore, if at least two of the fractions in the Farey triplet $[\frac{p_L}{q_L}, \frac{p_c}{q_C}, \frac{p_R}{q_R}]$ 
 that characterize butterflies
are friendly fractions, the determinant $D$ is unity and trace of $T$ is an integer. This implies that the scaling exponent $\zeta$ is an irrational number of the form  $\zeta= [n^*+1; \overline{ 1, n^*}]$ and therefore both type-I and type-II scaling belongs to same universality class.
However, the Eq. (\ref{t}) includes the possible scenario where that none of the pairs of fraction in the Farey triplet $[\frac{p_L}{q_L}, \frac{p_c}{q_C}, \frac{p_R}{q_R}]$ are friendly fractions and transformation maps two fractions with $D \ne 1$. This will lead to new universality classes, different from the class that describes Type-I butterflies. Whether the NNN model described here supports these new classes of butterflies has not been seen in our limited exploration of the parameter space. \\

(2) Missing in these recursions is the fact that it does not give the renormalization of $\Delta \sigma$. This is due to the fact that the transformation is defined on the Farey tree and not on the Wannier diagram that requires putting Farey tree on a torus.

 \subsection{ Anomalous Bands}
 
The energy spectrum of the generalized Harper model  hosts type-I,  type-II and type-III butterflies. This raises the natural question:
 which bands of the energy spectrum transform from type-I to  type-II  or type-III and which bands remain unchanged as NNN coupling $t_{ab}$ is tuned. In Harper model,
 there are  bands with ambiguous Chern Numbers as the Chern number to the left and the right of the band are not same as is the case for
 the central band at $\phi=1/3$. Two of the butterflies that share this band are $[2/7, 3/10, 1/3$]  ( $ N= 4$) and $[1/3, 3/8,2/5$ ( $N=-2$ ).  The two possible values $4$ and $-2$ are two possible solutions of the Diophantine equation $ qM+pN=1$ as it has infinity of solutions
 $ N = N_0 + n q$ as $ 4 = -2+2.3$ with $N_0=-2$ and $n= 2$.
 Such bands appear to transform in the presence of NNN terms, transforming type-I butterfly to  to type-II.
 

 \section{ Summary and Conclusions}
 
Wannier diagram, also known as the gap labeling theory  encodes some of the quintessential features of the energy spectrum of Bloch electrons.  Discovered soon after the discovery of the butterfly graph,  this alternative elegant description provided  an important benchmark for laboratory realization of the butterfly spectrum. In this paper, the Farey tree - a beautiful part of number theory,  is shown to to be intimately related to the Wannier diagram.  Consequently, the butterfly graph can be viewed as an incarnation of an abstract mathematical set that organizes all rationals between $0$ and $1$,
adding an immense simplicity and mystique to remarkable complexity of Bloch electrons in a magnetic field involving interplay between two competing periodicities. The central to this rather intriguing and non-intuitive simplicity lurking in the butterfly graph is  based on the key observation that the process of constructing Farey tree involves straight lines that have integer slopes and integer intercepts. Alternatively,  submerged in the  Wannier diagram - a graph of straight lines with integer slopes and integer intercepts, is a hierarchical lattice of trapezoids whose  parallel lines are  perpendicular to the $x$-axis, representing two fractions that are neighbors in the Farey tree.  Stated more explicitly,  given two vertical lines drawn at fractions $\frac{p_L}{q_L}$ and $\frac{p_R}{q_R}$ to the $x$-axis, of lengths $\frac{1}{q_L}$ and $\frac{1}{q_R}$ respectively,  where $p_Lq_R-p_Rq_L = \pm 1$, create trapezoids whose diagonals and slanting lines have integer slopes and integer intercepts.  Constructing a hierarchy of this lattice using Farey sum rule and stacking such trapezoids symmetrically in a unit square generates the entire Wannier diagram.\\
 
 Dwelling mostly on the number theoretical aspects of the butterfly spectrum, here we also unveil a simple rule that captures a non-number-theoretical characteristic.
Our observation that not all Wannier trajectories find representation in the  butterfly graph can be stated as  a simple rule of minimal violation of symmetry of the butterflies. In other words, quantum mechanics 
 of Bloch electrons in a magnetic field improvises on number theory,  to generate the butterfly spectrum. In a generalized Harper model, the Farey hierarchy prevails. Intriguingly, NNN model uses NNN Farey fractions
 to create  new species of butterflies. These butterflies with somewhat different number theoretical characteristics, are not the exact replica of the main butterfly. However, their recursions  can be described 
 by the renormalization group framework that describes the recursive structure of the Harper model.
 What perturbations take us outside this renormalization and perhaps outside the number theoretical description inherent in the energy spectrum remains an interesting open question.\\ 
 
 Our brief discussion of the butterfly graph in NNN model  explores a very small part of the multi-dimensional parameter space of the NNN model. Although the new species of butterflies were found to be described by the universality class of the Type-I butterflies, that is, are characterized by scaling exponent
 $\zeta= [1+n^*; \overline{1,n^*}]$, the  renormalization equations as described in Appendix reveal the possibility of new university classes.\\ 
 
As a final comment, we note that  missing in the recursions  described by Eq. (\ref{t}), is the fact that it does not give the renormalization of $\Delta \sigma$ - the symmetry parameter that uniquely labels all the butterfly of in the butterfly graph, in a given Farey triplet $[\frac{p_L}{q_L}, \frac{p_c}{q_c}, \frac{p_R}{q_R}]$. This requires renormalization equations for the trapezoidal cells. Our attempts for such renormalization fail to give elegant equations somewhat like Farey tree renormalization. The fact that Wannier trajectories have integer slopes and intercepts suggest that the recursions for these cells should have simplicity and  elegance  that we have not been able to demonstrate.

 \appendix
 
 \section{ Farey Tree Hierarchy and  M\"{o}bius Transformation}
 
We describe an  important symmetry property of the
Farey tree  where by the word  ``symmetry" , we do not refers to the Euclidean  geometrical symmetry, but symmetry described by
invertible algebraic transformations that maps one pair of Farey fractions to another.  That is, we seek a
transformation $T$ :
 
 \begin{equation}
  \Big(\frac{p_x(1)}{q_x(1)}, \frac{p_y(1)}{q_y(1)}\Big) \rightarrow  \Big(\frac{p_x(2)}{q_x(2)}, \frac{p_y(2)}{q_y(2)}\Big) = T  \Big(\frac{p_x(1)}{q_x(1)}, \frac{p_y(1)}{q_y(1)}\Big),
  \end{equation} 
 where each pair satisfies $ \Big(p_x(l)q_y(l)-p_y(l)q_x(l)\Big) =  D \ne 0$ ($ l=1,2$)  and  the mapping preserves the order, that is $\frac{p_x(1)}{q_x(1)} \rightarrow \frac{p_x(2)}{q_x(2)}$ and $\frac{p_y(1)}{q_y(1)} \rightarrow \frac{p_y(2)}{q_y(2)}$. To obtain $T$, we construct  two matrices $T_1$ and $T_2$ as:
 \begin{equation}
T_1 =   \begin{bmatrix} p_x(1) & p_y(1) \\ \\ q_x(1) & q_y(1) \end{bmatrix},\,\,\,\ T_2=  \begin{bmatrix} p_x(2) & p_y(2) \\ \\ q_x(2) & q_y(2) \end{bmatrix}
\label{t12}
\end{equation}

We will now show that  required map is\cite{Hatcher}:
\begin{equation}
 T  = T_2 T^{-1}_1
  =    \frac{1}{D}  \begin{bmatrix} p_x(2) q_y(1)-p_y(2)q_x(1) & p_x(1)p_y(2)-p_x(2)p_y(1) \\  \\ q_x(2)q_y(1)-q_x(1)q_y(2) & p_x(1) q_y(2)-p_y(1)q_x(2)  \end{bmatrix} 
 \label{t}
 \end{equation}

To prove Eq. (\ref{t}), consider a transformation that maps  a primitive fraction $\frac{p}{q}$ to another primitive fraction $\frac{p^{\prime}}{q^{\prime}}$, defined as,
 
 \begin{equation}
 \frac{p}{q} \rightarrow \frac{p^{\prime}}{q^{\prime}} = \frac{a p+ b q}{c p + d q} \equiv \frac{ a\frac{p}{q} + b} { c \frac{p}{q} + d},
 \label{T}
 \end{equation}
 
 The above equation can also be written as,
\begin{equation}
  \left( \begin{array}{cc} p \\ q \end{array} \right) \rightarrow \left( \begin{array}{cc} p^{\prime} \\ q^{\prime} \end{array} \right) =  \left( \begin{array}{cc} a & b \\ c & d 
\end{array} \right) \left( \begin{array}{cc} p \\ q \end{array} \right) \equiv \mathcal{M}  \left( \begin{array}{cc} p \\ q \end{array} \right)
\end{equation}

Under this transformation, $\frac{0}{1} \rightarrow \frac{b}{d}$ and $\frac{1}{0} \rightarrow \frac{a}{c}$. In other words, $\mathcal{M}^{-1}$ maps a pair of fractions $(\frac{b}{d}, \frac{a}{c} )$ to  $( \frac{0}{1}, \frac{1}{1})$.\\

Therefore, the transformation that maps $ \Big(\frac{p_x(1)}{q_x(1)}, \frac{p_y(1)}{q_y(1)}\Big)$ to   $ \Big(\frac{p_x(2)}{q_x(2)}, \frac{p_y(2)}{q_y(2)} \Big)$ can be constructed as a two step process where we first
 map $ \Big(\frac{p_x(1)}{q_x(1)}, \frac{p_y(1)}{q_y(1)}\Big)$
 to $( \frac{0}{1}, \frac{1}{1})$  where $\left( \begin{array}{cc} a & b \\ c & d 
\end{array} \right) = T_1^{-1}$ and then map $( \frac{0}{1}, \frac{1}{1})$ to $ \Big(\frac{p_x(2)}{q_x(2)}, \frac{p_y(2)}{q_y(2)}\Big)$ where $\left( \begin{array}{cc} a & b \\ c & d 
\end{array} \right) = T_2$.

This completes the proof that $T = T_2 T_1^{-1}$ where $(T_1, T_2)$ are given by equation  (\ref{t12}).\\
\\

For self-similar hierarchical structure,  the renormalization equation connecting two consecutive levels $l$ and $l+1$ is given by,
\begin{equation}
  \left( \begin{array}{cc} p(l+1) \\ q(l+1) \end{array} \right) = T   \left( \begin{array}{cc} p(l) \\ q(l) \end{array} \right)
\end{equation}
This equation  encoding the Farey tree hierarchy  also describe the recursive structure of the type-I, type-II  and the  type-III butterflies.
The eigenvalues of $T$ determine the asymptotic scalings of the butterfly flux interval.These eigenvalues are of the form $(\zeta, \zeta^{-1})$. This is because the matrix $T$ has real trace and its
determinant  is unity as from  the product rule of the determinant, $ Det[T]= Det[T_2].Det[T_1^{-1}] =D. \frac{1}{D} = 1$. \\
\\

 For type-I butterflies,  every sub-butterfly  is a renormalization of the main butterfly\cite{MW1, SW}.   In this recursive scheme,
every friendly triplet $[\frac{p_L}{q_L}, \frac{p_c}{q_c}, \frac{p_R}{q_R} ]$ is related to $[\frac{0}{1}, \frac{1}{2}, \frac{1}{1}]$ by a conformal map - a M\"{o}bius transformation. 
This transformation can be constructed by choosing a pair of friendly fractions.
For example,  we can choose $\frac{p_x(1)}{q_x(1)} = \frac{0}{1}$ and $\frac{p_y(1)}{q_y(1)} = \frac{1}{1}$ and we  write $\frac{p_x(2)}{q_x(2) } = \frac{p^*_L}{q_L^*}$ and $\frac{p_R(2)}{q_R(2) } = \frac{p^*_R}{q_R^*}$ and obtain a simplified recursion\cite{SW,Sat21},

 \begin{equation}
 \phi(l+1) = \frac{(p^*_R-p^*_L) \phi(l)+ p_L^*} {(q^*_R-q^*_L)  \phi(l) + q_L^*} \equiv \begin{bmatrix} (p^*_R-p^*_L)  &  p_L^* \\ (q^*_R-q^*_L) & q_L^* \end{bmatrix} 
  \begin{bmatrix} p(l)\\ q(l) \end{bmatrix}
 \label{RR}
 \end{equation}
 
 The transformation  also maps $\frac{1}{2}$ to $\phi_c$ as with $\phi(l)=\frac{1}{2}$, we get  $\phi(l+1) =  \frac{p^*_L+p^*_R}{q^*_L+q^*_R}$\
 
The eigenvalues of the transformation matrix, denoted as $(\zeta, \zeta^{-1})$,  determines  the asymptotic scaling factors and are given by:

\begin{equation}
\zeta= \lim _{ l \rightarrow \infty}  \frac{p_x(l+1)}{p_x(l)} = \lim _{ l \rightarrow \infty}  \frac{q_x(l+1)}{q_x(l)}  =\frac{( q^*_L+ p^*_R-p^*_L)}{2}\pm\sqrt{\left(\frac{q^*_ L+  p^*_R-p^*_L}{2}\right)^2-1} ,
 \label{zeta1}
 \end{equation}
Expressed as a  continued fraction expansion, these quadratic irrationals  are given by,
 \begin{equation}
 \zeta= [n^*+1; \overline{ 1, n^*}], \,\ n^* = q^*_L + p^*_R-p^*_L-2,
 \label{star}
 \end{equation}
 
 where
 \begin{eqnarray}
   [n^*+1; \overline{ 1, n^*}]  \equiv n^*+1+ \cfrac{1}{1
          + \cfrac{1}{n^*
          +\cfrac{1}{1
          +\cfrac{1}{n^*
          + \cfrac{1}{1 + \cfrac{1}{n^*.....} } } }}}
          \label{cont}
          \label{z}
          \end{eqnarray}
          
The scaling exponent (\ref{z})  describes type-I, type-II and type-III butterflies. The possibility of a new universality class requires  $|p_x q_y - p_y q_x| = D > 1$ for all the three pairs of fractions
in a given Farey triplet that defines a butterfly.


\begin{thebibliography}{99}
\bibitem{Hof}
D.R. Hofstadter,
{\it Phys.Rev.B}, {\bf 14}, (1976) 2239-49.
 \bibitem{book} I. I. Satija, {\it Butterfly in the Quantum World} IOP Concise,
Morgan and Claypool, San Raffael, CA, 2016.
\bibitem{PT} V. Galitski , G. Juzeliunas  and I.B. Spielman Phys Today, 72, 1, 38,  2019.
\bibitem{Math1}Avila A and Jitomirskaya S Ann. Math. 170 303 (2009).
\bibitem{Dean}C. R. Dean, L. Wang, P. Maher, C. Forsythe, F. Ghahari, Y. Gao, J. Katoch, M. Ishigami, P. Moon, M. Koshino, T. Taniguchi, K. Watanabe, K. L. Shepard, J. Hone and P. Kim 
Nature {\bf 497}, 598–602 (2013)
\bibitem{TKNN} D.J. Thouless ,  M. Kohmoto , M. P. Nightingale  and M. den Nijs,  Phys. Rev. Lett. {\bf 49}  ( 1982) 405.
\bibitem{Sat16}[11] I. I. Satija,
Eur. Phys. J. - Special Topics, 225, 2533-47 (2016).
\bibitem{SW} I. Satija and M. Wilkinson,  {\it J. Phys A} ,  {\bf 53},  085703, 2020.
\bibitem{Sat21} I.I. Satija, J. Phys. A: Math. Theor. 54 ( 2021), 025701
\bibitem{Harper} P.G. Harper,  Proc.
Phys. Soc. A 68  ( 1955) 874
\bibitem{NNN} F. Claro, Phys. Status Solidi (B) 104, K31 (1981).
\bibitem{NNNT} D. J. Thouless, Phys. Rev. B 28, 4272 (1983).
\bibitem{NNN1} Y. Hatsugai and M. Kohmoto, Phys. Rev. B 42, 8282 (1990).
\bibitem{NNN2}  J. H. Han, D. J. Thouless, H. Hiramoto, and M. Kohmoto, Phys.
Rev. B 50, 11365 (1994).
\bibitem{Math2} Avila A, Jitomirskaya S and Marx C A , 
 Math. 210 283, (2017) 
 \bibitem{MW1} M. Wilkinson,  J. Phys. A: Math. Gen. 20 4337 ( 1987).
\bibitem{MW2} M. Wilkinson, Proc. Roy. Soc. A 391 305–50 (1984).
\bibitem{W} G. H. Wannier, Phys. Status Solidi B  {\bf 88} , 757 (1978).
\bibitem{CW} F. H. Claro and G.H. Wannier, Phys Rev B {\bf 19} (1979) 6068.
\bibitem{Thouless} D. J. Thouless, Phys Rev B {\bf 27} 6083 (1983).
\bibitem{T83} D. J. Thouless, B {\bf 28} (1983) 4272.
\bibitem{JM} R. Johnson and J. Moser, Comm. Math. Phys, {\bf 84} (1982) 403;  {\bf 90} (1983) 317 (err).
\bibitem{DS} F. Delyon and B. Souillard, Comm. Math. Phys. {\bf 89} (1983) 415.
\bibitem{Simon} B. Simon, Adv. Appl. Math. {\bf 3} (1982) 463.
\bibitem{Mac} A. H. MacDonald , Phys. Rev. B {\bf 28}  6713 (1983);
Phys.Rev.B {\bf 29}  3057 (1984).
\bibitem{Dana} I. Danna, Y Avron and J. Zak, J. Phys C {\bf 18} L679 (1985).
\bibitem {Hatcher}  ``Topology of Numbers by Allen Hatcher, ebook, published 2018.

\end{thebibliography}
\end{document}